\documentstyle[prl,aps,twocolumn,epsf]{revtex}
\begin{document}
\draft

\twocolumn[\hsize\textwidth\columnwidth\hsize\csname
@twocolumnfalse\endcsname

\title{
Finger behavior of a shear thinning fluid
in a Hele-Shaw cell}
\author{Eugenia Corvera~Poir\'e and Martine Ben~Amar}
\address{
Laboratoire de Physique Statistique\\
\'Ecole Normale Sup\'erieure\\
24, rue Lhomond; 75231 Paris Cedex 05; France\\
e-mail: eugenia@lps.ens.fr; benamar@lps.ens.fr\\
}
\maketitle

\begin{abstract}
We make a theoretical study of the behavior of a simple fluid
displacing a shear thinning fluid confined in a Hele-Shaw cell. 
To study the Saffman-Taylor instability
when the displaced fluid is non Newtonian we face the problem of
having a field which is non laplacian.  By means of an hodographic
transformation we are able to solve the problem in the case of
weak shear thinning while taking into account the non laplacian character
of the equation.  Our results predict that the finger width
decreases towards zero for small values of the surface tension parameter 
which inversely proportional to the finger velocity.
\end{abstract}

\vskip2pc]

\narrowtext

Instabilities in complex fluids have been recently
the subject of intense research due to their importance in
technology. The classical hydrodynamic experiments, such as
the Dean flow and Taylor-Couette flow have been
considered to study bulk instabilities~\cite{shaqfeh}. 
On the other hand, interfacial instabilities between complex fluids,
with some exceptions~\cite{homsy2}, have been less studied.
Theoretically, this is due to the difficulty
of the equations governing these systems.
Here we consider the Saffman-Taylor problem for a viscous fluid
whose viscosity changes with shear. 
The Saffman-Taylor problem~\cite{bensimon,saffman2}
is the prediction of the steady-state
shape of the fluid interface in a two-phase flow confined in a 
linear Hele-Shaw cell.
A viscous fluid is pushed by a low viscosity fluid;
modes grow and compete dynamically, and the competition leads to a single 
finger-shaped pattern at large times~\cite{dynamics2}.
Experimentally~\cite{saffman2,experiment}, 
the finger is characterized by its width $\lambda$.
It is found that $\lambda > 1/2$ in units of the
channel width $W$, and that it is a unique function
of a control parameter $\sigma$ such that 
$\lambda \rightarrow \frac{1}{2}$ as $\sigma \rightarrow 0$. 
The surface-tension parameter $\sigma$ is defined as
\begin{equation}
\label{gamma}
\sigma \equiv \frac{4 T}{12 \mu U}
\left( \frac{b}{W} \right) ^2 \ \ \ \ ,
\end{equation}
where
$T$ is the surface tension, $\mu$ is the viscosity of the 
fluid that is being pushed, $b$ is the gap spacing of the Hele-Shaw cell
and $U$ is the velocity of the finger in the laboratory frame.
The control parameter $\sigma$ provides a singular perturbation 
to the zero-surface-tension equations of motion describing the
Hele-Shaw flow~\cite{solvability3}.
Several experiments were
carried out to alter this parameter and study the system's 
response~\cite{benamar,jacob,couder,maher1,chicago}.
New steady-state patterns were observed both in experiments and simulations
in these ``perturbed'' Hele-Shaw cells~\cite{couder,guo1,chicago}.
Most of this patterns were theoretically explained, by altering the
boundary conditions of the problem~\cite{benamar,aniso}
while keeping the Laplacian character of the equations (see below). 
Other patterns were explained by imposing
mathematical conditions at the finger tip~\cite{daniel}.
Recently, experiments have been made in which the fluid which is pushed
is non Newtonian. Some of these experiments have used
clays~\cite{vandamme2} and others polymer solutions~\cite{maher3}.
Both groups report seeing a fracture-like behavior of the
complex fluid. 
Non-Newtonian fluids differ widely in their physical properties,
with different fluids exhibiting a range of different effects
from plasticity and elasticity to shear thickening and shear thinning.
In this paper we are interested in studying how the viscous fingering
regime is altered when considering
an effect such as shear thinning or shear thickening,
{\it i.e.}, we consider a non-Newtonian fluid whose viscosity depends on
the local shearing.

Classically the constant-velocity displacement of a viscous fluid
by a fluid of negligible viscosity such as air is governed by Darcy's law,
which states that the velocity is proportional to the pressure gradient.
Recently a generalized Darcy's law where viscosity depends upon the
squared pressure gradient has been derived by using a fluid model
with shear rate dependent viscosity \cite{shelley}. As pointed out by them,
this can be formally inverted to give a generalized Darcy's law where
viscosity depends on the magnitude of the local velocity. We consider
the latter for the present work.
For a constant viscosity fluid, Darcy's law plus the incompressibility of
fluids lead to a Laplace's equation for the pressure. This allows one
to use well established conformal mapping techniques~\cite{mclean}.
For a non-constant viscosity fluid, imposing the incompressibility 
of fluids does not lead to Laplace's equation for the pressure.
This poses a serious problem since one is unable to apply standard techniques.
In the laboratory frame, where ${\vec v}_{plates}=0$,
the equations governing our system are the generalized Darcy's law
where viscosity depends on the magnitude of the local two-dimensional 
fluid velocity and the incompressibility of fluids:
\begin{equation}
\label{fund}
{\vec v} = -\frac{b^2}{12 \mu (V)} \nabla p \,\,\,\, , \,\,\,\,\,\,\,\,\,\,\,\,\,\,\
\,\,\,\,\,\,\,\,\,\,\,\, 
\nabla\cdot{\vec v} = 0 \,\,\,\, .
\end{equation}
With the following boundary conditions
\begin{equation}
\label{bcs}
\left.  \begin{array}{l} v_n=U\sin\theta_T \\ p_0-p=\frac{T}{R} \end{array} \right\} \mbox{on the finger}  \,\,\,\, ,
\end{equation}
$v_y=0$ on the walls,
$v_x=V_\infty=U\lambda$, $v_y=0$ as $x\rightarrow\infty$ for $-1 < y < 1$
$v_x=0$, $v_y=0$ \,\,\,\, $x\rightarrow -\infty$ for  $\lambda < |y| < 1$.
Here $R$ is the local radius of curvature, $V$ is the magnitude of the
local fluid velocity in the laboratory frame of reference,
$\theta_T$ is the angle tangent to the finger measured from the $x$-axis
and the subfix $n$ stands for the component of the velocity
normal to the interface.  We have set $W=2$. 
It is worth it to explicitly point out that
equation \ref{fund} does not violate Galilean invariance since it is
only valid in the laboratory frame of reference.
We are going to treat problems for which viscosity
depends on velocity, 
so we write the viscosity as
$\frac{1}{\mu}=\frac{1}{\mu_0} (\frac{\rho}{\rho_0})$,
where $\rho$ depends on the velocity and $\rho_0$ is the
function $\rho$ evaluated at the finger tip, $\mu_0$ is a constant
defined in such a way as to be the viscosity of the fluid when there
is no dependence on the velocity.
Defining $\phi \equiv \frac{b^2}{12 \mu_0} p$, equations~\ref{fund}
can be written in a differential form as
\begin{equation}
\label{darcy}
d\phi=\frac{\rho_0}{\rho}(v_x dx+v_y dy) \,\,\,\, ,
\end{equation}
and 
\begin{equation}
\label{inc}
d\chi=-v_y dx+v_x dy \,\,\,\, .
\end{equation}
We write the complex velocity in polar coordinates 
as $(v_x-i v_y) = V \exp{-i\theta}$
and express $dz=dx+i dy$ as
\begin{equation}
dz=\frac{\exp{i\theta}}{V}(\frac{\rho}{\rho_0} d\phi+i d\chi) \,\,\,\, .
\end{equation}
The basic problem that we face is the non-Laplacian character
of the equation describing the problem, {\it i.e.}, equations \ref{fund} do
not lead to Laplace's equation for the pressure.

We make an hodographic transformation by exchanging dependent and
independent variables. So we consider $x(V,\theta)$ and $y(V,\theta)$.
We express $d\phi$ and $d\chi$ in terms of partial derivatives
as functions of $(V,\theta)$. Since $dz$ is an exact differential
its crossed derivatives should be equal. This gives an
equation whose real and imaginary parts provide identities among
the partial derivatives of $\chi$ and $\phi$ with respect to $V$ and $\theta$.
Such identities allow us to write 
equations for $\chi(V,\theta)$ and $\phi(V,\theta)$ by equating
the crossed derivatives obtained by differentiating them.

We work out the case in which the viscosity depends on the
velocity as a power law, {\it i.e.}, $\rho=V^\alpha$. 
This is one of the standard forms for the viscosity found
in literature. In this case, the equations for $\chi(V,\theta)$ 
and $\phi(V,\theta)$
have the following form
\begin{equation}
\label{eqnchixt}
\frac{\partial^2 \chi}{\partial V^2}+
\frac{(1-\alpha)}{V}\frac{\partial \chi}{\partial V}+
\frac{1}{V^2}
\frac{\partial^2 \chi}{\partial \tau^2}=0 \,\,\,\, ,
\end{equation}
and
\begin{equation}
\label{eqnpxt}
\frac{\partial^2 \phi}{\partial V^2}+
\frac{(1+\alpha)}{V}\frac{\partial \phi}{\partial V}+
\frac{1}{V^2}
\frac{\partial^2 \phi}{\partial \tau^2}=0 \,\,\,\, .
\end{equation}
Where the angle has been redefined as {$\tau\equiv (1-\alpha)^{-1/2}\theta$}.
We rescale the fields $\chi$ and $\phi$ in such a way as to obtain
Laplace equation to linear order in $\alpha$,
{\it i.e.},
for the fields
\begin{equation}
{\cal P}=V^{\alpha/2}\phi \,\,\,\, \mbox{and} \,\,\,\,{\cal X}=V^{-\alpha/2}\chi \,\,\,\, ,
\end{equation}
we have the following equations
\begin{equation}
\label{eqncalx}
\frac{\partial^2 {\cal X}}{\partial V^2}+
\frac{1}{V}\frac{\partial {\cal X}}{\partial V}+
\frac{1}{V^2}
\frac{\partial^2 {\cal X}}{\partial \tau^2}=
\frac{\alpha^2}{4}\frac{{\cal X}}{V^2} \,\,\,\, ,
\end{equation}
and
\begin{equation}
\label{eqncalp}
\frac{\partial^2 {\cal P}}{\partial V^2}+
\frac{1}{V}\frac{\partial {\cal P}}{\partial V}+
\frac{1}{V^2}
\frac{\partial^2 {\cal P}}{\partial \tau^2}=
\frac{\alpha^2}{4}\frac{{\cal P}}{V^2} \,\,\,\, .
\end{equation}
Up to this point we have shown that, for the viscosity depending
on a power of the velocity
\begin{equation}
\label{Laplacians}
\nabla^2_V {\cal P} = {\cal O}(\alpha^2)  \,\,\,\, \mbox{and} \,\,\,\, \\
\nabla^2_V {\cal X} = {\cal O}(\alpha^2)  \,\,\,\, ,
\end{equation}
where $\nabla^2_V$ is the Laplacian operator in the velocity plane.

 The hodograph method and the rescaling of fields, despite being little
intuitive from a physical point of view, have permitted us to
write the equations of the problem in such a way as to allow for an
approximation in a controlled manner. Now, for small $\alpha$ which
corresponds to the case of weak shear thinning, we approximate
${\cal P}$ and ${\cal X}$ by laplacian fields. This leads 
to the following expression for $dz$
\begin{equation}
\label{dzeq}
dz=\frac{\exp{i\tau(1-\frac{\alpha}{2})}}{V^{(1-\frac{\alpha}{2})}}(d{\cal P}+i d{\cal X}) \,\,\,\, .
\end{equation}
When writing the differentials for ${\cal P}$ and ${\cal X}$ from \ref{dzeq},
it is easy to see that, if we define an effective velocity of the form
${\vec v^*}=v^*_x\hat{\imath} + v^*_y\hat{\jmath}$ with
$v^*_x= V^{(1-\frac{\alpha}{2})}\cos\tau(1-\frac{\alpha}{2})$ and
$v^*_y= V^{(1-\frac{\alpha}{2})}\sin\tau(1-\frac{\alpha}{2})$
we have equations that are mathematically equivalent to classical
Darcy's law and the incompressibility of fluids, {\it i.e.},
\begin{equation}
{\vec v^*} = \nabla{\cal P} \,\,\,\, , \,\,\,\,\,\,\,\,\,\,\,\,\,\,\
\,\,\,\,\,\,\,\,\,\,\,\,
\nabla\cdot{\vec v^*} = 0 \,\,\,\, .
\end{equation}
We emphasize here that ${\vec v^*}$ is not a real velocity;
and that ${\cal P}$ is not a pressure, but a function of pressure
and the local fluid velocity. Nothing new has been introduced
by defining ${\vec v^*}$. 
We have found equations which are mathematically equivalent to the classical
problem, and therefore, we are now dealing with Laplacian 
fields in real space.
We are now in the position of performing conformal transformations.
We work in a frame of reference in which
\begin{equation}
\label{phipsi}
\Phi=\frac{{\cal P}-\Im{\cal H}}{\omega} \,\,\,\, \mbox{and} \,\,\,\,
\Psi=\frac{{\cal X}+\Re{\cal H}}{\omega} \,\,\,\, ,
\end{equation}
where ${\cal H}$ is an unknown analytic function whose values at the contour
are set so that the contour (wall+finger+center line) maps
into the real axis of the $(s,t)$ plane of the following conformal
transformation
\begin{equation}
\label{map}
\sigma=s+it=\exp{-[\Phi-\Phi_0+i\Psi+i{\cal H}]\pi} \,\,\,\, .
\end{equation}
This method has been used previously in reference~\cite{benamar3}.
McLean and Saffman have solved the problem of constant viscosity for
arbitrary surface tension~\cite{mclean}. In the rest of the analytical treatment
we follow closely their technique.
Details of the derivation will be presented in a longer publication.
In the variables $\theta=\theta_T-\pi$ and $q_N=q(1-\lambda)$
we find
\begin{eqnarray}
\kappa sq_N\frac{d}{ds} \left[V^{\alpha/2}sq_N\frac{d\theta}{ds}\right]
 -q_N=  \nonumber \\
 -\frac{q_N}{1-\lambda} -\frac{q_N U}{\pi {\cal Q}_0/\lambda}\int^{1}_{0}
\frac{V^{-\alpha/2}\sin\theta}{q_N(s'-s)}ds' \label{first} \,\,\,\, ,
\end{eqnarray}
where
\begin{equation}
\kappa\equiv\frac{Tb^2\pi^2}{12\mu_0U^\alpha(1-\lambda)^2{\cal Q}_0/\lambda}
\,\,\,\, '
\end{equation}
$q\equiv\frac{d\Phi}{dS}$ where $dS$ means differenciation along the interface
and ${\cal Q}_0=V_\infty^{1-\alpha/2}$.
Equation \ref{first} reduces to McLean-Saffman equation when $\alpha=0$.
A second equation is obtained by analyticity properties and 
it is exactly like the McLean and Saffman relation among $q_N$ and $\theta$,
{\it i.e.}, 
\begin{equation}
\label{second}
\log q_N(s)= -\frac{s}{\pi}\int^1_0 \frac{\theta(s')}{s'(s'-s)}ds'
\,\,\,\, .
\end{equation}
Note that in this case $q_N$
no longer represents the finger velocity at the finger-tip frame.
The third equation is obtained by writing explicitly the fact that
the square of the velocity is equal to the sum of the squares of the
normal and tangential components, and it is
\begin{eqnarray}
\label{third}
\kappa sq_N\frac{d}{ds}
\left[V^{\alpha/2}sq_N\frac{d\theta}{ds}\right] = \nonumber \\
\frac{V^{1-\alpha/2}}{{\cal Q}_0/\lambda} 
\sqrt{1-\left(\frac{U}{V}\right)^2\sin^2\theta} \,\,\,\, .
\end{eqnarray}

\begin{figure}[h]
\centerline{\epsfxsize=220pt\epsfbox{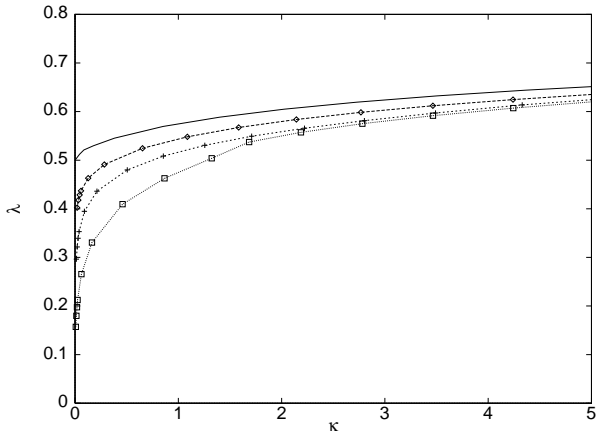}}
\caption
{ Finger width as a function of the surface tension
parameter $\kappa$. The solid line is the line of constant
viscosity ($\alpha=0$), the other lines correspond to the case of shear
thinning with exponents: $\diamond$ $\alpha=0.2$, $+$  $\alpha=0.4$,
$\sqcap$ $\alpha=0.6$.}
\label{fig1}
\end{figure}

Equations \ref{first}, \ref{second}, and \ref{third} generalize the
Saffman-Taylor problem to the case in which viscosity depends
as a power law on the local velocity.
They are solved numerically
with the method described in \cite{mclean}. 
In figure~\ref{fig1} we plot finger width as a function of $\kappa$
for the case of shear thinning ($\alpha>0$) which is the case more
often seen in experiments. We also plot for reference the case $\alpha=0$,
for which we obviously recover the McLean-Saffman solution to the
problem since the viscosity is a constant. 
At large values of $\kappa$ 
the solutions are very close to the solutions of McLean-Saffman. On the other
hand, at small values of $\kappa$ the solutions decrease dramatically
towards a zero finger width.
Between these two regions there is an intermediate region in which 
solutions go smoothly from one type of behaviour to the other one.
In all cases, solutions deviate from the $\alpha=0$ solution and the
deviation is stronger the larger the value of $\alpha$. 
We present results for $\alpha=0.2,0.4,0.6$
in order to illustrate how the solution changes with the power $\alpha$.
We have also performed a solvability analysis in the case of shear thinning.
Such analysis gives a scaling among the finger width $\lambda$
and the surface tension $\sigma$ in the
small $\sigma$ limit. The result is
\begin{equation}
\lambda\sim \sigma^{\alpha/((1+\alpha)(2-\alpha))}
\,\,\,\, .
\end{equation}
It is unfortunate that numerically it is impossible to verify
such a scaling, since it would imply computations at values of $\sigma$
out of our numerical capabilities. Therefore, this result should be
taken as complementary to our numerics.

We have treated the problem of viscous fingers in the presence of shear
thinning. We have obtained fingers whose width decreases towards zero
as the surface tension parameter goes to zero.
A detailed derivation of the equations presented here,
comparison with the $\eta$-model~\cite{benamar2} and 
a description of the solvability analysis will be presented 
in a longer publication. In this forthcoming publication we will
present as well results for shear thickening which, surprisingly,
present a dramatic decrease of finger width at very small $\kappa$.
Also, since our model is two dimensional,
it is valid as long as the finger width is larger than the height
of the cell.
Effects due to a third dimension, which might be very
important, specially in connection to fracture behavior,
have not been considered in the present work.

\end{document}